\documentclass[twocolumn,showpacs,amsmath,amssymb,superscriptaddress,nofootinbib]{revtex4}

\usepackage{graphicx}
\usepackage{epsfig}
\usepackage{dcolumn}
\usepackage{bm}

\newcommand{\beq}{\begin{equation}}
\newcommand{\eeq}{\end{equation}}
\newcommand{\bea}{\begin{eqnarray}}
\newcommand{\eea}{\end{eqnarray}}

\begin{document}
\title{Triaxial nuclear models and the outer crust of nonaccreting cold neutron stars}
\author{Lu Guo, M. Hempel, J. Schaffner-Bielich and J. A. Maruhn}
\affiliation {Institut f{\"u}r Theoretische Physik, J. W. Goethe-Universit{\"a}t, 60438 Frankfurt am Main, Germany}
\date{\today}

\begin{abstract}
The properties and composition of the outer crust of nonaccreting
cold neutron stars are studied by applying the model of Baym,
Pethick, and Sutherland (BPS) and taking into account for the first
time triaxial deformations of nuclei. Two theoretical nuclear
models, Hartree-Fock plus pairing in the BCS approximation (HF-BCS)
with Skyrme SLy6 parametrization and Hartree-Fock-Bogolyubov (HFB)
with Gogny D1S force, are used to calculate the nuclear masses. The
two theoretical calculations are compared concerning their neutron
drip line, binding energies, magic neutron numbers, and the sequence
of nuclei in the outer crust of nonaccreting cold neutron stars,
with special emphasis on the effect of triaxial deformations. The
BPS model is extended by the higher-order corrections for the atomic
binding, screening, exchange and zero-point energies. The influence
of the higher-order corrections on the sequence of the outer crust
is investigated.
\end{abstract}


\maketitle

\section{\label{level1} Introduction}
Neutron stars are believed to be born as the central compact remnant of an ordinary star which
exploded in a core-collapse supernova. With initial temperatures above $10^{11}$ K one can assume
that the matter of the neutron star is at local nuclear statistical equilibrium. For a nonaccreting
neutron star it is plausible that its composition is determined by this condition even when the
neutron star has cooled down to much lower temperature (here we are ignoring effects from possible
contaminations of fall-back supernova material).

Within the first few centimeters of the outermost layers of a cooled-down neutron star the
density already becomes so high that atoms begin to touch. At $\rho \sim 10^4$ g/cm$^3$ the atoms are
completely ionized and separated into their constituents, the electrons and nuclei. The light
electrons form a degenerate Fermi gas, soon becoming relativistic (at $\rho \sim 10^7$ g/cm$^3$),
whereas the massive nuclei arrange in a solid (bcc)-lattice to minimize the Coulomb-interaction. To
determine the ground state nucleus in the sequence of the outer crust of nonaccreting neutron stars,
 it is essential to know the binding energies of all nuclei from
the valley of $\beta$-stability up to the neutron drip line. With the lowest energy per
nucleon, $^{56}$Fe is present at low density. With increasing baryon and electron chemical potential (whereas the latter is fixed by charge neutrality) the ground-state nuclei become
more and more neutron-rich until finally no more neutrons can be bound. This happens at the neutron
drip ($\rho_{ND}\sim 4.3\times10^{11}$ g/cm$^3$), where free neutrons begin to appear and the inner
crust begins.

In 1971 Baym, Pethick and Sutherland (BPS) \cite{BPS} analyzed this
form of cold dense matter and calculated the resulting sequence of
nuclei and the equation of state of the outer crust of nonaccreting
cold neutron stars. They used the nuclear data from the droplet
model of Myers and Swiatecki \cite{MyersSwiatecki} for their
analysis. Over the years more and more masses of unstable nuclei
were measured experimentally. In parallel the theoretical nuclear
models, needed for nuclei with unknown mass, have been developed
further. Several publications followed, which updated the results of
BPS by using the latest experimental and theoretical nuclear data:
In 1989 Haensel, Zdunik and Dobaczewski \cite{HZD} used a Skyrme
Hartree-Fock-Bogolyubov (HFB) calculation with the parameter set SkP
in spherical approximation \cite{HFBHZD}, and a newer version of the
droplet model from Myers \cite{Myers}. Haensel and Pichon \cite{HP}
in 1994 used the experimental nuclear data from the 1992 atomic mass
table of Audi and Wapstra \cite{Audi93} and the theoretical nuclear
mass tables of the droplet models from M\"oller and Nix
\cite{Moeller}, as well as the Skyrme parametrization of Aboussir et
al.~\cite{Sksc4}. The last update was done by R\"uster, Hempel and
Schaffner-Bielich \cite{RHS} by applying the 2003 atomic mass table
from Audi, Wapstra, and Thibault \cite{AW} and a comprehensive set
of 21 different theoretical nuclear models. Besides the finite-range
droplet model (FRDM) \cite{Moller95, Moller97} various
non-relativistic Skyrme-parameterizations and mass tables based on
relativistic nuclear field theories were employed. Furthermore the
effects of pairing and axial deformations were examined. In the
present work, for the first time the impact of triaxial deformations
on the neutron drip line and the sequence of nuclei of the outer
crust will be studied. Haensel and Pichon \cite{HP} extended the
original BPS model by including higher-order corrections of the
atomic binding, screening, and exchange energies. In the present
calculation these corrections are also applied and their influence
on the equation of state and the sequence of ground state nuclei
will be analyzed.

The paper is organized as follows. Section~\ref{level2} gives a brief outline of the two triaxial nuclear models. The BPS model and its higher-order corrections are present in Sec.~\ref{level3}. The binding energies, the location of drip lines, and the sequences of the outer crust between the two theoretical models are compared in Sec.~\ref{level4}. The effect of higer order corrections is also discussed. Section~\ref{level5} is devoted to the summary and conclusion.

\section{\label{level2} Description of the triaxial nuclear models}

In this section, we describe the two mean-field methods \cite{Bender03} used in the present studies:
a Skyrme energy functional with BCS pairing and a HFB theory with the finite-range Gogny interaction.  Both
calculations include nuclear triaxial features, where the gamma degree of freedom plays
an important role in the description of deformed nuclei.

\subsection{Skyrme Hartree-Fock plus BCS}\label{skyrmepar}

Density functional theory states that the total energy of many-body system can be formulated by an
energy functional
\beq
\mathcal{E} = \mathcal{E}_{\textrm {kin}}[\tau] +  \mathcal{E}_{\textrm {Sk}}[\rho,\tau,\vec{J}]  +
                        \mathcal{E}_{\textrm {C}}[\rho_p] - \mathcal{E}_{\textrm {c.m.}},
\eeq
which only depends on the local distribution of density $\rho$, kinetic energy density $\tau$ and spin-orbit
current $\vec{J}$. Here $\mathcal{E}_{\textrm {kin}}$ is the functional of the kinetic energy,
$\mathcal{E}_{\textrm {Sk}}$ the Skyrme energy functional, $\mathcal{E}_{\textrm {C}}$
the Coulomb functional including the exchange term in Slater approximation, and $\mathcal{E}_{\textrm
{c.m.}}$ the center-of-mass correction \cite{Bender99,Rein06}.
The pairing correlations are treated in the BCS approximation using a delta pairing force
\cite{Tondeur83,Krieger90},
$V_{\textrm{pair}}(\vec{r},\vec{{r}^\prime})=V_q\delta(\vec{r}-\vec{{r}^\prime})$. The pairing
strength $V_p$ for the protons and $V_n$ for the neutrons are fitted to the pairing gaps in isotopic
and isotonic chains \cite{Fle04b}.
The pairing energy functional is given by
\beq
\mathcal{E}_{\textrm {pair}} = \frac{1}{4} \sum_{q={p,n}}V_q \int d^3r \Delta_q^2,
\eeq
with $\Delta_q$ the pairing density.

The variation of the energy functional $\mathcal{E}$ with respect to the single-particle wave
functions yields the mean-field equations
\beq
\hat h\phi_k = \epsilon_k\phi_k  \hspace{0.5cm} \textrm {with}  \hspace{0.5cm}
\hat h = \frac{\partial\mathcal{E}}{\partial\hat\rho}.
\eeq
The coupled HF-BCS
equations are solved on a grid in coordinate space with the damped gradient iteration method
\cite{Blum92} and a Fourier representation of the derivatives. No symmetry restriction has been
imposed in the calculation.

In the present work, we have chosen the SLy6 parametrization
\cite{Chabanat98} from recent fits. The fitting of the force SLy6
laid particular emphasis on the properties of neutron matter and
neutron rich nuclei in order to improve the isospin properties away
from the $\beta$-stability line. Nuclear matter properties, the
binding energies and radii of the doubly closed-shell nuclei
$^{40,48}$Ca, $^{56}$Ni, $^{132}$Sn, and $^{208}$Pb were used for
the fit.  The SLy6 force is a good candidate for describing isotopic
properties of nuclei from the $\beta$-stability line to the drip
line.

\subsection{Gogny Hartree-Fock-Bogoliubov}
The other method used in the present study is HFB with finite-range Gogny interaction
\cite{Gog80,Egido93,Guo04}.  The HFB equation
\beq
\left(\begin{array}{cc} h & \Delta \\ -\Delta^\ast & -h^\ast \end{array} \right)
\left(\begin{array}{cc} U_k \\  V_k \end{array} \right) = \epsilon_k \left(\begin{array}{cc} U_k \\
V_k \end{array} \right),
\eeq
is expressed in terms of $h = t +\Gamma-\lambda$ with $\Gamma$ being the HF potential and $\Delta$
the pairing potential. The total energy of the nuclear system $E_0$ reads
\beq
E_0 = {\textrm {Tr}}(t\rho)+\frac{1}{2}{\textrm {Tr}}(\Gamma\rho)
-\frac{1}{2}{\textrm {Tr}}(\Delta\kappa ^\ast),
\eeq
with the density matrix $\rho$ and pairing tensor $\kappa$.

The HFB equation was solved in a three-dimensional
harmonic-oscillator basis \cite{Guo04,Guo05,Guo06}. The triaxial
oscillator parameters in the Hermite polynomials were optimized for
each nucleus to maximize the ground-state binding energy.  All the
contributions to the HF and pairing fields arising from the Gogny
and Coulomb interactions as well as the two-body correction of the
kinetic energy are included in the self-consistent procedure. Here
we employed the finite-range Gogny force with parametrization D1S
\cite{Gog84, Gog91}, which was adjusted to give a better description
of the nuclear surface energy.  The finite-range Gogny force
provides both the HF mean-field and pairing field simultaneously in
the framework of the full HFB theory.

\section{\label{level3}The extended BPS model}
To calculate the outer crust of nonaccreting cold neutron stars with the two triaxial nuclear
models, we start from the BPS model \cite{BPS} specified in \cite{RHS}. The total energy density
\begin{equation}
\epsilon_{tot}=n_N(W_N+W_L)+\epsilon_e \; ,
\label{etot}
\end{equation}
describes completely ionized nuclei with mass $W_N$, and number density $n_N$ immersed within an ideal Fermi gas of electrons ($\epsilon_e$) and the corresponding Coulomb interaction. The nuclei are arranged in a
body-centered cubic lattice, represented by the lattice energy $W_L$. In the present work the energy density Eq.~(\ref{etot}) is extended with the higher-order corrections of atomic binding ($B_{el}$), screening ($W_{Sc}$), exchange ($W_{Ex}$), and zero-point energy ($W{zp}$),
\begin{equation}
\epsilon_{tot}=n_N(W_N+W_L+B_{el}+W_{Sc}+W_{Ex}+W{zp})+\epsilon_e \; .
\end{equation}

For the masses of the nuclei $W_N$, which are the essential input information for the calculation, we
always prefer taking the experimental data of the atomic mass table 2003 from Audi, Wapstra, and
Thibault \cite{AW}, if available. As nuclear binding energies are needed, the tabulated values are
corrected for the included atomic electron binding energy with an empiric formula given in
\cite{LPT}
\begin{equation}
B_{el}=\left(14.4381Z^{2.39}+1.55468\cdot 10^{-6}Z^{5.35}\right) \mbox{eV} \; ,
\label{bel}
\end{equation}
with the charge number $Z$. The predicted theoretical binding energies do not include any atomic interaction, hence this correction is not necessary for them.

The screening or Thomas-Fermi energy \cite{Salpeter}
\begin{eqnarray}
W_{Sc} & = & -\frac{162}{175}\left( \frac{4}{9\pi}\right)^{2/3}\alpha^2Z^{7/3}\mu_e
\end{eqnarray}
includes deviations of the electron distribution from uniformity caused by the positively charged ions,
where $\mu_e$ is the electron chemical potential and $\alpha$ the fine structure constant.

The third correction, the exchange energy~\cite{Salpeter}
\begin{eqnarray}
W_{Ex} &=& -Z \frac3{4 \pi}\alpha\, m_e c^2 \, x \, \phi(x) \label{wex} \; ,
\end{eqnarray}
where
\begin{eqnarray}
x & = & k_{F_e}/m_ec
\end{eqnarray}
and
\begin{eqnarray}
\phi(x) & = &\frac{1}{4x^4}\left[\frac94+ 3\left(\beta^2 -\frac{1}{\beta^2}\right) \ln\beta - 6\left(\ln\beta\right)^2\right. \nonumber \\
& & -\left.\left(\beta^2+ \frac{1}{\beta^2}\right)-\frac18\left(\beta^4+\frac1{\beta^4}\right)\right]
\end{eqnarray}
with $\beta=x+\sqrt{1+x^2}$ and the mass $m_e$ and the Fermi momentum $k_{F_e}$ of the electrons,
takes the fermionic nature of the electron-electron interaction into account.

The zero-point motion of the nuclei in the lattice \cite{Salpeter}
\begin{equation}
W_{zp} = \frac32 \hbar \sqrt{\frac{4 \pi \alpha \hbar c Z^2n_N }{3 W_N}}
\end{equation}
is much smaller
than the first three corrections, and is only included to verify the stability of the lattice.
Its influence on the composition and the equation of state is negligible.

\begin{figure}
\centering
\includegraphics[width=8.6cm]{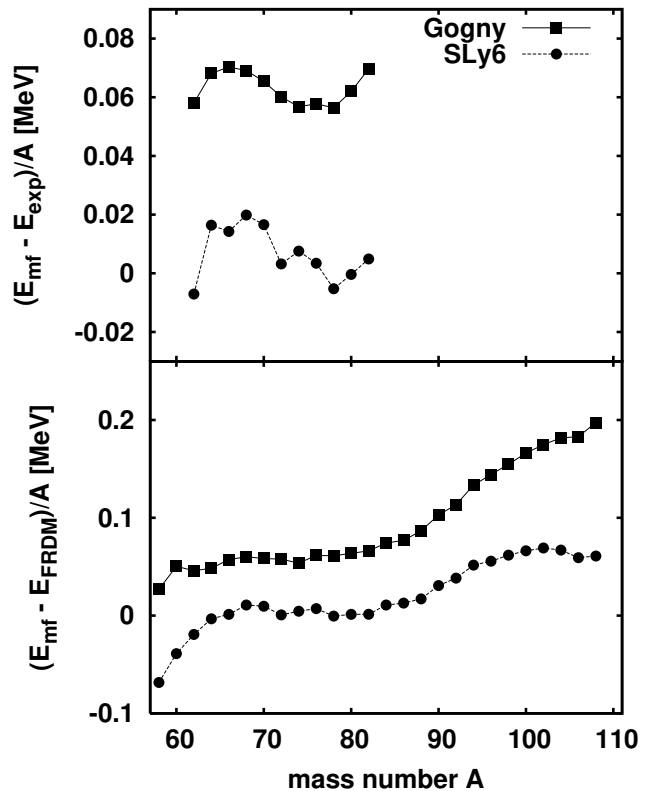}
\caption{Binding energies per nucleon of Ge isotopes from the proton drip-line to the neutron drip-line
with Gogny and SLy6 forces.
The lower panel shows the energy differences to FRDM values
\cite{Moller95}
and the upper panel the differences to the experimental data
\cite{Audi03}.}
\label{figa}
\end{figure}

\begin{figure}
\centering
\includegraphics[width=8.6cm]{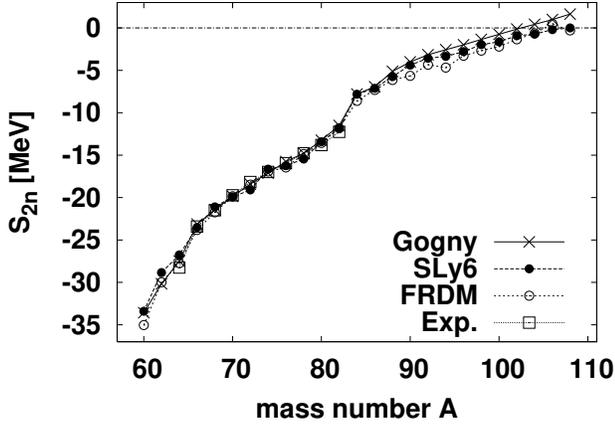}
\caption{Two neutron separation energies for Gogny, SLy6, FRDM, and the available experimental data.}
\label{figb}
\end{figure}

\section{\label{level4}Discussion and Results}
We systematically calculate the properties of ground states from
stable to neutron-rich nuclei with proton number $Z=20$ to 50 using
two triaxial nuclear models, HFB with Gogny D1S parametrization and
HF-BCS with Skyrme SLy6 parametrization. Both calculations predicted
that most of Germanium isotopes are triaxially deformed, where the
calculated quadrupole and triaxial deformations are in good
agreement with the available experimental data \cite{Guo07}.
Triaxiality is also predicted for some other nuclei with small
differences between the two theoretical nuclear models. As an
example, we select the typical triaxial deformed Ge isotopes and
compare the calculated energies of two theoretical models with the
available experimental data and the finite-range droplet model
(FRDM). Figure \ref{figa} shows the energies per nucleon of the Ge
isotopes from the proton-rich isotope with neutron number $N=26$ to
neutron-rich nuclei up to the neutron number $N=76$ which is beyond
the neutron drip-line of all sets discussed here (see below). Owing
to a huge energy variation along the long isotopic chain, we show
the relative energies per nucleon, in the upper panel relative to
the experimental data and in the lower panel relative to the results
from the macroscopic-microscopic model FRDM taken from
\cite{Moller95}. In the range of experimentally measured isotopes,
Gogny HFB calculation underestimates the binding energy throughout.
The deviation of the results for SLy6 HF-BCS from the experimental
data is small which is no surprise because SLy6 is a more recent
parametrization. The lower part of the figure shows the relative
energies in a much broader range of isotopes with respect to the
FRDM as experimental data is not available in the neutron-rich
regime. The trends seen in the range of experimentally accessible
nuclei are basically continued for neutron rich nuclei. The Gogny
HFB calculation tends to lower binding energies while the SLy6
HF-BCS shows relatively small deviations in comparison to the FRDM.

Figure \ref{figb} shows the two neutron separation energies for Gogny HFB, SLy6 HF-BCS, FRDM,
and the available experimental data. The overall agreement of the two-neutron
separation energies is fair
among different models and the experimental data, especially in the
experimentally measured regime up to $A=82$,
although Gogny HFB calculation systematically underestimates the binding energies.
The location of drip line from the Gogny HFB calculation is different from that of
SLy6 and FRDM, e.g. Gogny HFB predicts
the location of drip-line at the nucleus $^{102}$Ge while the SLy6 HF-BCS and FRDM
predicts $^{106}$Ge as being the last stable nucleus.

As shown in \cite{RHS}, the equation of state of the outer crust of
nonaccreting cold neutron stars is almost independent of the
theoretical nuclear model used, and this remains valid for the
presently examined triaxial models. The transitions from one
equilibrium nucleus to another happen at different densities, while
the overall form of the equation of state does not change. The
higher-order corrections affect the equation of state only very
slightly. Electronic versions of the equation of state and the
chemical composition for the model SLy6 HF-BCS and some other
selected nuclear models described in \cite{RHS}, can be downloaded
as tables from one of the authors' webpages
\footnote{\texttt{http://www.th.physik.uni-frankfurt.de/$\sim$hempel}}
or the EPAPS online archive \footnote{See EPAPS Document No.
E-PRVCAN-76-004801 for electronic versions of the equation of state
and chemical composition in tabular form for six different selected
nuclear models. For more information on EPAPS, see
http://www.aip.org/pubservs/epaps.html.}. All of these tables
include the higher order corrections.

\begin{figure}
\centering
\includegraphics[width=8.6cm]{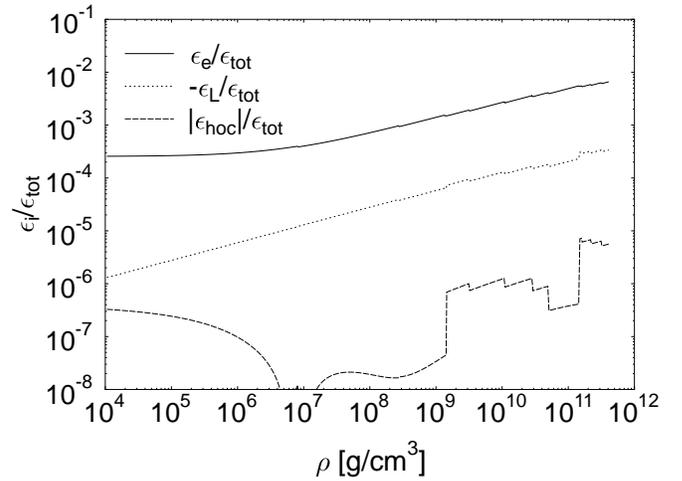}
\caption{Fractions of the energy density of the electrons $\epsilon_e$, the lattice $\epsilon_L$, and the sum of the higher-order corrections $\epsilon_{hoc}$ on the total energy density $\epsilon_{tot}$, calculated with the theoretical nuclear model SLy6 HF-BCS.}
\label{fig1}
\end{figure}

Figure \ref{fig1} shows the different contributions to the total energy density, which is dominated by the mass of the nuclei. The energy density of the electrons is two to three orders of magnitude smaller, followed by the negative lattice energy density $\epsilon_L=W_L n_N$, which is again one to two orders of magnitude below the electron contribution. The sum of the higher-order corrections $\epsilon_{hoc}=n_N(B_{el}+W_{Sc}+W_{Ex}+W_{zp})$ is positive for densities below $10^7$ g/cm$^3$ and negative for higher mass densities. The highest fraction of the order of $10^{-6}$ to $10^{-5}$ on the total energy density is reached at high densities above $10^9$ g/cm$^3$. The jumps in the curves arise from the transitions of one ground state nucleus to another.

\begin{figure}
\centering
\includegraphics[width=8.6cm]{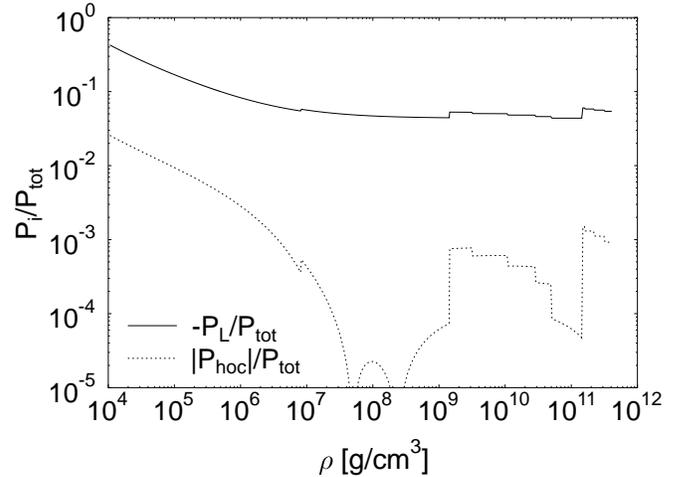}
\caption{Fractions of the pressure of the lattice $P_L$, and the sum of the higher-order corrections $P_{hoc}$ on the total pressure $P_{tot}$, calculated with the theoretical nuclear model SLy6 HF-BCS.}
\label{fig2}
\end{figure}

In Fig.~\ref{fig2} the contributions to the total pressure are depicted. The pressure is mainly generated by the electrons. The lattice gives a significant contribution of about 40\% at low densities which decreases down to about 4\% at high densities. The pressure of the higher-order corrections is always negative, besides the little bump around $10^8$ g/cm$^3$. The biggest influence of the higher-order corrections on the equation of state is seen at smaller densities: they lower the pressure up to $\sim 3\%$. Above $\rho \sim 10^6$ g/cm$^3$ their contribution drops below $10^{-3}$. At densities larger than $10^9$ g/cm$^3$ the contribution raises again to fractions of $10^{-4}$ to $10^{-3}$.

\begin{figure}
\centering
\includegraphics[width=8.6cm]{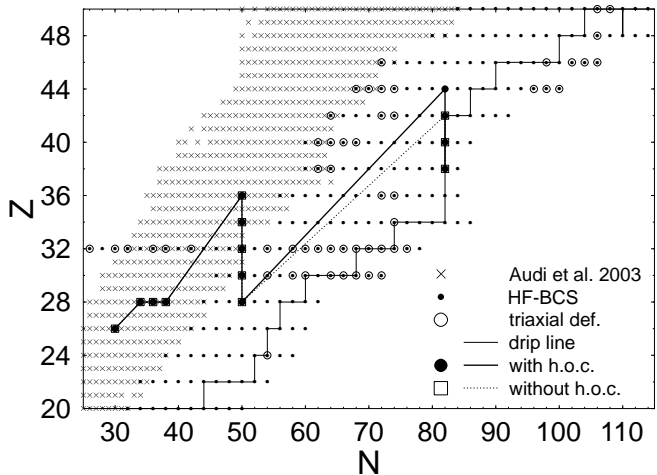}
\caption{Nuclear chart (proton number $Z$, neutron number $N$) of the nuclei considered in this paper, taken from the atomic mass table~\cite{AW}
(crosses) and calculated results with the SLy6 HF-BCS model (dots). The open circles mark triaxially
deformed nuclei. The thick line with full circles shows the
sequence of ground state nuclei in the outer crust of nonaccreting cold neutron stars with increasing
density, the thin line the neutron drip line of the theoretical nuclear model. The dashed line with squares represents the sequence without the higher-order corrections.}
\label{fig3}
\end{figure}

\begin{figure}
\centering
\includegraphics[width=8.6cm]{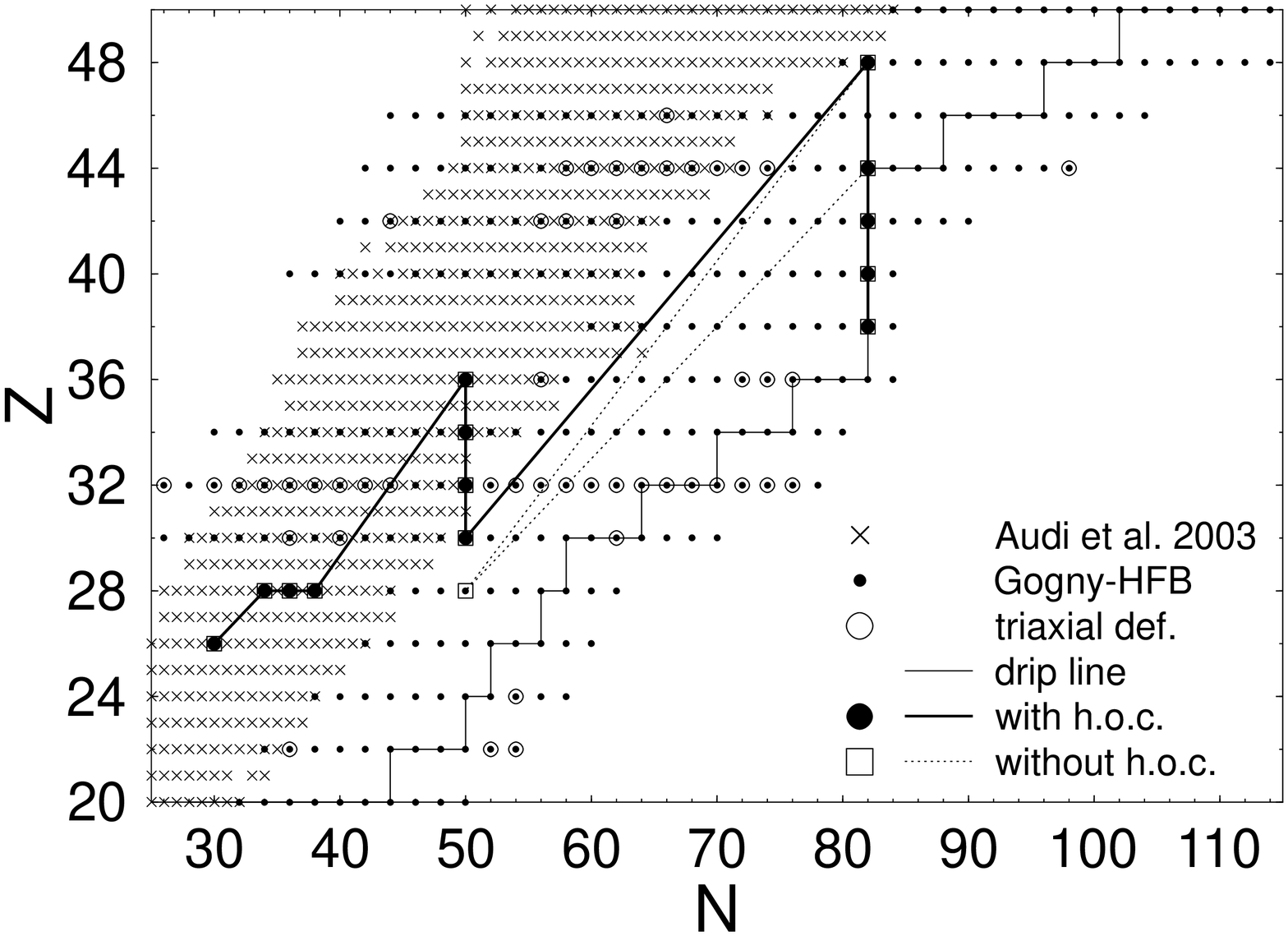}
\caption{As Fig.~\ref{fig3}, now for the Gogny HFB calculation.}
\label{fig4}
\end{figure}

Figures \ref{fig3} and \ref{fig4} show all the nuclei examined in this work, using the atomic mass
table and the two theoretical models studied here. Nuclei in the theoretical models which fulfill
the criteria $0.03<\beta$ and $5^\circ<\gamma<55^\circ$ for the deformation parameters $\beta$ and
$\gamma$ are marked additionally with an open circle as being "triaxially deformed".
In both models almost
all the Germanium-isotopes ($Z=32$) are triaxially deformed.
Triaxial nuclei are also found in the region of $56\le N \le74$ and $38\le Z \le 46$. The Gogny HFB model shows only a few other triaxial nuclei, while for the SLy6 HF-BCS model there are some more at $Z=30$ and in the region of $Z\ge 44 $, and $96\le N \le108$.

The drip lines of the two theoretical calculations as determined via
the two-neutron separation energy are also depicted. For both models
one recognizes a rather linear increase at small $N$, followed by a
pronounced vertical step at the magic neutron number $N=82$. In
comparison to the non-relativistic "state-of-the-art" theoretical
models in \cite{RHS}, e.g. BSk8, SLy4, and FRDM, the drip line of
the SLy6 HF-BCS model shows a similar behavior up to $N=82$. For
larger neutron numbers its drip line is located at $\Delta Z=0-4$
larger values of proton number $Z$. The drip line of the triaxial
HF-BCS calculation is also similar to the axial Skyrme calculations
\cite{Stoitsov03}, and is only shifted by one or two nuclei for some
isotopes. If one compares the drip line of SLy6 HF-BCS to the Gogny
HFB, one sees that the Gogny HFB drip line is shifted about $\Delta
N=0-6$ to smaller neutron numbers.

The sequence of nuclei which form the ground state in the outer crust of nonaccreting neutron stars
with increasing density is much more sensitive to the underlying theoretical nuclear model than the
equation of state. Figures \ref{fig3} and \ref{fig4} show the results for the sequences of
the two triaxial calculations. The last nucleus of the sequence with experimentally determined
mass is $^{80}$Zn which both models have in common. For the HF-BCS model at densities above $4.9\times 10^{10}$ g/cm$^3$ the theoretically calculated nucleus $^{78}$Ni with the neutron magic number
50 appears in the sequence.
 Afterwards the sequence continues along the pronounced neutron magic number 82 from
$^{126}_{~44}$Ru to $^{120}_{~38}$Sr, where the neutron drip line is reached. Compared to
the previous calculations in \cite{RHS}, the sequence of the triaxial HF-BCS model is the same as that of
FRDM \cite{Moller95, Moller97}, except that the sequence of HF-BCS is
only shifted by $\Delta Z=2$ to nuclei with larger $Z$ at $N=82$.

The Gogny HFB calculation shows some distinctive features in comparison to the triaxial HF-BCS model
and the
theoretical nuclear models studied in \cite{RHS}. After $^{80}$Zn the sequence jumps to
$^{130}_{~48}$Cd, as seen in Fig.~\ref{fig4}, which still originates from the atomic mass table \cite{AW}. None of the theoretical
models examined in \cite{RHS} predicted such a high proton number nucleus in the sequence. Afterwards a gap in the $N=82$ isotone chain appears, until the first experimentally unknown nucleus $^{126}_{~44}$Ru shows up in the sequence. The following $N=82$ sequence is the same as in the triaxial HF-BCS calculation.

This behavior may be attributed to the fact that in the analyzed region the nuclei of the Gogny HFB
calculation are more weakly bound compared to the nuclei with measured mass and to the other theoretical models. Also the location of the drip line enhances the view that the Gogny HFB calculation predicts relatively low binding energies for neutron-rich nuclei. The parameterization D1S was constructed with emphasis on the surface energy
and not particularly for describing binding energies of neutron rich nuclei, which might be one of
the reasons for the observed deviations from the other nuclear sequences. But even more important
could be that the calculation was performed in the harmonic oscillator basis.

In order to show the effect of higher order corrections,
Figures \ref{fig3} and \ref{fig4} depict the sequences where the higher-order corrections were neglected. As expected, the higher-order corrections only lead to small changes in the sequence, which occur at high densities ($\rho>1.4\times 10^{11}$ g/cm$^3$). For the HF-BCS model the nucleus $^{126}_{~44}$Ru drops out in the sequence, while for Gogny HFB $^{78}$Ni enters in after $^{130}_{~48}$Cd. The attraction caused by the higher-order corrections slightly favors nuclei with larger proton and mass number. Without the higher-order corrections, as done in \cite{RHS}, the sequence of the HF-BCS model is more similar to the one of the FRDM \cite{RHS}. The Gogny HFB calculation shows the uncommon features, namely the appearance of large proton numbers and the jump in the sequence.

None of the nuclei appearing in the outer crust is triaxially (and not even axially)
deformed. Deformations become important in the regions far from closed shells, whereas the sequence
runs mainly along the neutron magic numbers. The effect of the additional degree of freedom of
triaxial deformation is small compared to the enhanced binding energy of magic nuclei, thus the
sequence shows no significant differences to most of the axially deformed or even spherically
calculated nuclei \cite{RHS}.

\section{\label{level5}Summary}
In the present work we have investigated for the first time the
impact of triaxial deformations on the neutron drip line and the
sequence of nuclei in the outer crust of nonaccreting cold neutron
stars. Two theoretical nuclear models HF-BCS and Gogny HFB,
including nuclear triaxial feature, have been used to systematically
study the ground state binding energies and triaxial deformations
for neutron-rich nuclei from proton number $Z$=20 to 50. These
neutron-rich nuclei will become experimentally accessible in the
near future with the ISAC-$\amalg$ and upcoming facilities as FAIR
at GSI Darmstadt. Taking the triaxially deformed Ge isotopes as an
example, we found that the Gogny HFB calculation systematically
underestimates the binding energies in comparison to the FRDM
results and to the experimental data, while the HF-BCS calculation
shows smaller deviations. The two-neutron separation energies among
Gogny HFB, SLy6 HF-BCS, FRDM and the experimental data are in good
agreement around the valley of stability, while the Gogny HFB
calculation predicts the drip-line nucleus to be $^{102}$Ge compared
to $^{106}$Ge for SLy6 HF-BCS and FRDM.

The location of the drip line of the triaxial HF-BCS calculation is
similar to that of the axial Skyrme calculation \cite{Stoitsov03},
and is shifted by one or two nuclei for some isotopes. In comparison
to the non-relativistic theoretical models in \cite{RHS}, e.g. BSk8,
SLy4, and FRDM, the drip line of the HF-BCS calculation shows a
similar behavior up to $N=82$, while for larger neutron numbers it
is located at somewhat larger values of $Z$. The drip line of the
Gogny HFB calculation lies at smaller neutron numbers compared to
the HF-BCS case, indicating overall slightly lower binding energies.
In both models the magic neutron number $N=82$ is observed as a
pronounced vertical step in the drip lines.

We presented the sequence of ground state nuclei for the outer crust
of neutron stars with an extended BPS model. The sequence of the
HF-BCS calculation is similar to those obtained in \cite{RHS} for
spherical or axially deformed nuclei, following mainly the magic
neutron numbers $N=50$ and 82. The sequence for the Gogny HFB
calculation exhibits deviations from the other nuclear models,
visibly by the appearance of the nucleus $^{130}_{~48}$Cd with a
large proton number and a jump to $^{126}_{~44}$Ru in the sequence
thereafter. The reason for this behavior is the comparatively low
binding energy of the nuclei in this region, which is probably due
to the calculation being performed in the harmonic oscillator basis
or to the parametrization D1S itself.

The composition of the outer crust of neutron stars and the role of
higher order corrections have been studied also in detail. The
biggest influence of the higher order correction on the equation of
state was found at lower densities, where the higher order
corrections decrease the total pressure by a few percent. For larger
densities the higher order corrections were almost negligible for
the equation of state. Nevertheless, they led to small changes in
the sequence of nuclei which appear at high densities
($\rho>10^{11}$ g/cm$^3$).

In both calculations no triaxially deformed nuclei were present in
the sequence of nuclei. One can conclude that the effect of triaxial
deformations is too weak to have significant consequences for the
sequence of nuclei of the outer crust of neutron stars.

In conclusion, the location of the drip line and the sequence of
nuclei in outer crust of cold nonaccreting neutron stars seem to be
rather robust predictions being nearly insensitive to the parameter
set, the approximation scheme and even to triaxial deformations for
most of the state-of-the-art nuclear models. These findings are of
importance for future rare isotope facilities as ISAC-$\amalg$ and
FAIR at GSI Darmstadt, where binding energies of neutron-rich nuclei
will be experimentally determined in the near future.

\begin{acknowledgments}
Lu Guo acknowledges support from the Alexander von Humboldt Foundation.
Matthias Hempel was supported from the Frankfurt Institute for Advanced Studies
and the Helmholtz Research School for Quark Matter Studies. The calculations were
performed at the Frankfurt Center for Scientific Computing which is gratefully acknowledged.
\end{acknowledgments}

\bibliography{astro}
\end{document}